\documentclass[10pt,superscriptaddress,prl,twocolumn]{revtex4}
\usepackage{amsfonts}
\usepackage{amsmath}
\usepackage{amssymb}
\usepackage{graphicx}
\usepackage{graphics}
\usepackage[usenames]{color}

\begin{document}
\title{Controlling ultracold chemical reactions via Rydberg-dressed interactions}
\author{Jia Wang}
\affiliation{Department of Physics, University of Connecticut, 2152
Hillside Rd., Storrs, CT 06269, USA}
\author{Jason N. Byrd}
\affiliation{Department of Physics, University of Connecticut, 2152
Hillside Rd., Storrs, CT 06269, USA}
\affiliation{Quantum Theory Project, University of Florida, Gainesville, FL
32611}
\author{Ion Simbotin}
\affiliation{Department of Physics, University of Connecticut, 2152
Hillside Rd., Storrs, CT 06269, USA}
\author{R. C\^ot\'e}
\affiliation{Department of Physics, University of Connecticut, 2152
Hillside Rd., Storrs, CT 06269, USA}
\affiliation{Institute for Quantum Computing, University of Waterloo, Waterloo, 
Ontario, Canada N2L 3G1}
\begin{abstract}

We show that ultracold chemical reactions can be manipulated and controlled 
by using Rydberg-dressed interactions. Scattering in the ultracold regime is sensitive to 
long-range interactions, especially when weakly bound (or quasi-bound) states exist
near the collision threshold. We investigate how, by Rydberg-dressing a reactant, one
enhances its polarizability and modifies the long-range van der Waals
collision complex, which can alter chemical reaction rates by shifting 
the position of near threshold bound states. We carry 
out a full quantum mechanical scattering calculation for the benchmark system 
H$_2$+D, and show that resonances can be moved substantially 
and that rate coefficients at cold and ultracold temperatures can be increased by 
several orders of magnitude.  

\end{abstract}

\maketitle

A key advantage of ultracold systems is the extraordinary degree of control 
they provide, such as tunable interactions through Feshbach resonances 
\cite{ChinReview2010} used to investigate degenerate quantum gases 
\cite{RMP-bose-1,RMP-bose-2,RMP-fermi}. This control 
allows to probe exotic three-body Efimov states \cite{efimov}, 
and to study ultracold molecules \cite{paper-JILA,sawyer2011} and modify 
their chemistry \cite{QuemenerReview2012,KremsIRPC2005,WeckIRPC2006}, 
{\it e.g.}, by orienting them \cite{paper-JILA,Jason-PRL}. Another approach to modify interactions
is to excite atoms into Rydberg states  \cite{HeidemannPRL2008},
where they acquire extreme properties ({\it e.g.}, long lifetimes or  
large electric dipole moment) \cite{gallagher94};
long-range Rydberg {\it trilobite} molecules 
\cite{GreenePRL2000,pfau} or {\it macrodimers} \cite{macrodimer-1,macrodimer-2} 
exemplify these exaggerated properties.
The control over strong interactions led to proposals for
quantum computing \cite{RMP-Saffman}, {\it e.g.}, to achieve quantum gates
\cite{JakschPRL2000,ProtsenkoPRA2002} or study quantum random walks \cite{CoteNJP2006}, and to the excitation blockade
mechanism \cite{blockade}, where a Rydberg atom prevents the excitation
of nearby atoms \cite{TongPRL2004,SingerPRL2004,VogtPRL2006,LiebischPRL2005,HeidemannPRL2008}; this effect is used to realize electromagnetically
induced transparency \cite{Rydberg-EIT-1,Rydberg-EIT-2},
to generate single photons \cite{lukin-single-photon} and photon-photon interactions
\cite{photon-photon1,photon-photon2}, 
or non-destructive imaging of Rydberg atoms \cite{matthias-imaging}
to study dynamics
of energy transport \cite{weidemuller-science}.

\begin{figure}[t]
\includegraphics[width=0.9\linewidth]{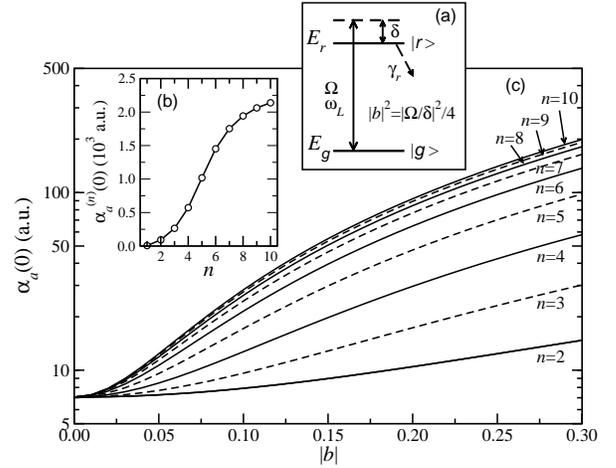}
\caption{(a) Sketch: the ground state $\left| g \right\rangle$ of energy $E_g$
              is dressed by a laser of frequency $\omega_L$ blue-detuned by $\delta$
              from a Rydberg level $\left| r \right\rangle$ of energy $E_r$ and  
              natural width $\gamma_r$ ($\Omega$ and $|b|^2$ are the
              Rabi frequency and fraction of Rydberg state, respectively).
              (b) Static polarizability of D excited into a state $np$ for $n\leq 10$. 
              (c) Static polarizability of D as a function of the Rydberg-dressing 
              coefficient $|b|$ (see text for details).}
\label{fig:sketch}
\end{figure}

Recent studies propose using Rydberg dressing 
to explore many-body physics \cite{ZollerPRL2010,HonerPRL2010} such as 
dipolar BEC \cite{ZollerPRL2000}, supersolid vortex crystals in 
BEC \cite{HenkelPRL2012} and to cool
polar molecules \cite{ZollerPRL2012}.
In this letter, we show how ultracold chemical reactions can be modified and controlled by 
Rydberg-dressing an atom approaching a diatom, which increases its polarizability
and modifies the atom-diatom van der Waals complex and the reaction rate. We consider H$_2$+D, 
a benchmark system for quantum calculations  
explored extensively at ultralow \cite{SimbotinPCCP2011} and higher 
temperatures \cite{h2-studies}, and for which accurate \emph{ab initio} 
potential energy surfaces (PES) \cite{BKMPJCP1996,MielkeJCP2002} exist.
Here, the deuterium atom D is Rydberg-dressed by weakly coupling its 
ground state $\left| g \right\rangle$ to a Rydberg state $\left| r \right\rangle$ 
of width $\gamma_r$ using a far detuned continuous-wave (CW) linearly polarized laser (see 
Fig.~\ref{fig:sketch}(a)). At large separation, when the atom-molecule
interaction is negligible, the atom can be modeled as a two-level system.
The CW laser, described by an oscillating electric field 
${\mathcal{E}}\cos \omega_L t$ of strength $\mathcal{E}$ and frequency 
$\omega_L$, couples $|g\rangle$ (energy $E_g$) and $|r\rangle$ (energy $E_r$)
with Rabi frequency $\hbar \Omega = \langle g | \mu \mathcal{E}| r \rangle$,
($\mu$:  dipole transition moment). The detuning $\delta$ is defined by $\hbar \delta=\hbar \omega_L - \left({ E_r-E_g}\right)$.

Under far-detuning ($|\delta|\gg\gamma_r$) and low laser power 
($|\Omega| \ll |\delta |$) conditions, the state of the atom 
(up to a global phase) within the rotating wave approximation (RWA) 
is $|a\rangle \approx \sqrt{1-|b|^2}|g\rangle + b|r\rangle$, 
where $b=\frac{1}{2}(\Omega/\delta ) e^{-i (\omega_L t - \phi_0)}$, and 
$\phi_0$ is an initial phase. Properties of dressed atom can be obtained using 
$|a\rangle$; {\it e.g.}, its dynamic polarizability $\alpha_a(\omega)$ 
and dipole moment $d_a$
\begin{equation}
   \alpha _a ( \omega ) = \left({1-|b|^2}\right)\alpha _a^{(g)}(\omega ) 
   + |b|^2 \alpha _a^{( r)}( \omega ),
   \label{alphaRyd}
\end{equation}
\begin{equation}
   d_a = \left({1-| b |^2}\right)d_a^{(g)}  + | b |^2 d_a^{(r)},
   \label{dipoleRyd}
\end{equation}
where $\alpha _a^{(j)}( \omega ) = \sum\limits_{k \ne j} 
\frac{2\omega _{jk} | \langle j |\mu | k \rangle |^2}
{\hbar ( \omega _{jk}^2  - \omega ^2 )}$
and $d_a^{(j)}=\langle j |\mu | j \rangle$ 
are the dynamic polarizability and dipole moment of the atom in state 
$| j \rangle$, respectively: $\hbar\omega_{jk}\equiv E_k-E_j$,
and the fast oscillating cross terms containing 
$\langle g |\mu | r \rangle$ are neglected. The static polarizability $\alpha _a^{(r)}(0)$ for Rydberg states is usually much 
larger than that of the ground state $\alpha _a^{(g)}(0)$ (since $\hbar\omega_{jk}$
can be small).A linearly polarized laser can excite D (or an alkali atom) 
from its atomic $s$ ground state into a Rydberg state $|j\rangle = |n,\ell_e,m_\ell\rangle$
(with principle quantum number $n$, electron angular momentum $\ell_e=1$
with projection $m_\ell =0$); Fig.~\ref{fig:sketch}(b)
shows the rapid growth of $\alpha _a^{(n)}(0)$ for the $np$ state of D ($n\leq 10$).
Note that without external fields mixing states of different parities, $d_a^{(j)}=0$ 
({\it e.g.}, for D in a pure $np$ state).
While $d_a^{(g)}$ remains small, even a weak electric field leads to Stark splittings 
of Rydberg states: in the linear Stark regime, the highest state of a given manifold 
$n$ has a large dipole moment $d_a^{(n)}=\frac{3}{2}n(n-1)$ a.u. \cite{ZollerPRL2010}. Even with a small mixing $|b|^2=|\Omega/\delta|^2/4$, a 
Rydberg-dressed atom can still possess a large polarizability 
(see Fig.~\ref{fig:sketch}(c)) and in some cases 
a large dipole moment which determine the long-range interaction 
with another atom or molecule. 

\begin{figure}[t]
\includegraphics[width=0.45 \textwidth]{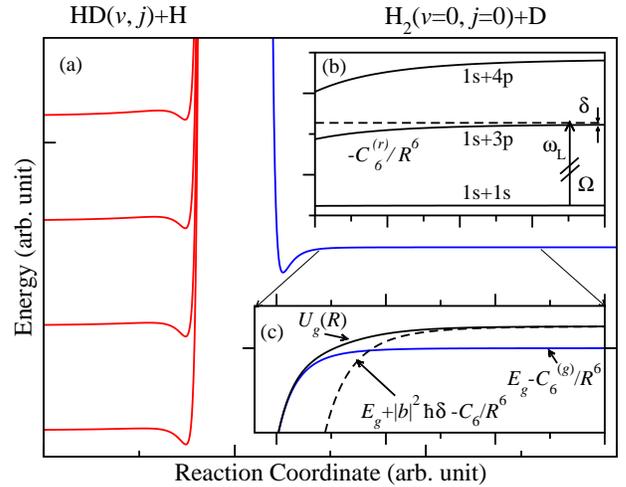}
\caption{(Color online) (a) Sketch of the PES vs. reaction coordinates (see text)
showing exit channels (left curves) nearest to the entrance channel on the right. (b) Rydberg dressing by a blue-detuned laser 
($\delta>0$) to avoid crossing a Rydberg curve. (c) Long-range interaction 
in the entrance channel: the dressed curve $U_g(R)$ (solid black line)
is more attractive with its threshold lifted from the {\it bare} case 
(blue solid line). $U_g(R)$ transitions smoothly from its asymptotic
form (dashed line) at large separation to the bare curve at shorter range.}
\label{fig:sketch-pot}
\end{figure}

In Fig.~\ref{fig:sketch-pot}(a), we sketch the PES's 
dependence on the reaction coordinates ({\it i.e.} the distance between H$_2$ and D 
in the entrance channel, and HD and H in the exit channels). Rydberg-dressing D changes the
long-range interaction in the entrance channel; for H$_2$
in its ground electronic state (no permanent dipole moment), the leading
interaction $-C_6/R^6$ depends on the van der Waals coefficient $C_6$.
Here, we assume the distance $R$ between D and H$_2$ is
large enough, and neglect the anisotropy of H$_2$. 
Note that if the molecule has a permanent dipole moment ({\it e.g.}, for heteronuclear 
molecules) and the atom is Rydberg dressed to a Stark state, 
the leading interaction is the dipole-dipole interaction.
Two components contribute to $C_6=C_6^{({\rm ind})}  + C_6^{({\rm dis})}$, 
a dipole induced-dipole term $C_6^{({\rm ind})}\propto d_a^2$, and a dispersion term
$C_6^{({\rm dis})}\propto \alpha_a$.
By changing $\alpha_a$ and $d_a$ using
Rydberg-dressing, one can modify the long-range atom-molecule van der Waals
complex, and affect scattering properties.

The dispersion term $C_6^{( j,{\rm dis})}$ for an atom in state 
$|j\rangle$ can be evaluated by the Casimir-Polder integral
\begin{equation}
   C_6^{( j,{\rm dis})}  = \frac{3}{\pi }\int_0^\infty  \alpha _m ( i\omega )
   \alpha _a^{(j)}( i\omega )d\omega  ,
\end{equation}
where $\alpha _m$ and $\alpha _a^{(j)}$ are the dynamical polarizability 
of the molecule and of the atom, respectively.
Inserting Eq.(\ref{alphaRyd}) into the Casimir-Polder integral gives
\begin{equation}
    C_6^{( {\rm dis})}  = C_6^{(g,{\rm dis})}  + | b |^2  \Delta C_6^{(r,{\rm dis})} ,
\end{equation}
where $ \Delta C_6^{(r,{\rm dis})} \equiv C_6^{( r,{\rm dis})}-C_6^{(g,{\rm dis})}$.The corresponding induction term is $C^{(j,{\rm ind})}_6 = \frac{2}{3} 
[d_a^{(j)} ]^2 \left(\alpha_{zz}-2\alpha_{xx}\right) \;$  \cite{ByrdJCP2011}, where $\alpha_{pq}$ is the $pq$ component of the cartesian static molecular dipole polarizability tensor (with $z$ along the molecular symmetry-axis). Inserting Eq. (\ref{dipoleRyd}) and noting that $d_a^{(g)}=0$, we find $C^{({\rm ind})}_6=|b|^4 C^{(r,{\rm ind})}_6 \;$.

The components of $C_6$ for the benchmark system H$_2$+D, 
tabulated in Table~\ref{tab1}, show a rapid increase with $n$. They
were computed using the TD-DFT/vdW method \cite{ByrdJCP2011} developed 
to obtain the dynamic polarizability for H$_2$ using the large aug-cc-pVTZ basis set \cite{DunningJCP1989} and the PBE0 density function.  The dynamic polarizability of D
was computed using Gauss-Laguerre quadrature, where the integration grid and 
number of excited states (computed using the proper reduced mass for D)
was converged to provide an accuracy better than 0.1\% (continuum contributions were 
omitted for D, suggesting a 5-10\% underestimate of $C_6^{(n,{\rm dis})}$). 
$C_6$ was computed using gaussian quadrature, with the induced term calculated 
for the highest Stark-splitted state for a given $n$.

\begin{table}
\caption{Calculated $C_6$ terms for a Rydberg-dressed D atom interacting 
with a ground state H$_2$ molecule: $[n]$ denotes $10^n$.} \label{tab1}
\begin{ruledtabular}
\begin{tabular}{c c c |  c c c }
$n$ & $C^{(n,{\rm dis})}_6$
  & $C^{(n,{\rm ind})}_6$ & $n$ & $C^{(n,{\rm dis})}_6$
   & $C^{(n,{\rm ind})}_6$\\
  & (a.u.)
  & (a.u.) & & (a.u.)
  & (a.u.)\\  
\hline
1  & 7.053   & 0.000 & 5  & 1.018 [3]   & 1.030 [4] \\
2  & 9.241 [1]   & 1.030 [2]  & 6  & 1.452 [3]   & 2.317 [4] \\
3  & 2.640 [2]   & 9.266 [2]  & 7  & 1.751 [3]   & 4.540 [4] \\
4  & 5.720 [2]   & 3.706 [3]  &8  & 1.942 [3]   & 8.072 [4] 
\end{tabular}
\end{ruledtabular}
\end{table}

In our benchmark example H$_2$+D, we
consider dressing D with a CW laser blue-detuned
$(\delta>0)$ from $3p$ (Fig.~\ref{fig:sketch-pot}(b)) 
to prevent populating the Rydberg state; 
as D approaches H$_2$, the detuning increases leading to
a weaker effect. For simplicity, we also consider no static field and omit
$C^{({\rm ind})}_6$; we can then drop the superscript (dis) and 
write $C_6=C_6^{(g)}+|b|^2 \Delta C_6^{(r)}$. Within the RWA, 
the Hamiltonian for the internal degrees of freedom at a fixed large 
separation $R$ can be written as a $2\times2$ matrix 
(in the basis $|am\rangle\equiv |a\rangle\otimes|m\rangle$ where
$|a\rangle = |g\rangle$ or $|r\rangle$ is the atomic state, and $|m\rangle$
the state of the molecule):
\begin{equation}
H = \left( {\begin{array}{*{20}c}
   - C_6^{(g)} /R^6  & \hbar \Omega/2   \\
   \hbar \Omega^*/2  & -\hbar \delta  - C_6^{(r)} /R^6   \\
\end{array}} \right),
\label{eq:Hamiltonian}
\end{equation}
where the ground state atom-molecule collision threshold is set to zero.
By diagonalizing (\ref{eq:Hamiltonian}), we obtain two 
Born-Oppenheimer curves: a dressed
ground curve $U_g$ between H$_2$ and ground-state D (dressed by $|r\rangle$),
and a dressed excited curve $U_r$ between H$_2$ and excited D (dressed by $|g\rangle$).
Assuming $| \Omega /\delta| \ll 1$ and $C_6^{\left( r \right)} > C_6^{\left( g \right)}$, 
we find
\begin{equation}\label{BO2b}
   U_g ( R ) =  - \frac{C_6^{(g)}}{R^6} + \Delta(R), \mbox{ with } 
   \Delta(R)\!\equiv\!\frac{\hbar|\Omega|^2}{4\delta( R )} \;,
\end{equation}
where $\hbar \delta ( R ) = \hbar \delta  + \Delta C_6^{( r )} /R^6$, 
with $\Delta C_6^{( r )}=C_6^{( r )}-C_6^{( g )}$. In the large $R$-limit such
that $\Delta C_6^{( r )} /R^6 \ll \hbar|\delta |$, the long-range 
behavior of $U_g(R) $ is given by
 \begin{equation}\label{BOlr}
   U_g ( R ) \to  - \frac{C_6^{( g )} \!\! + \!| b |^2 \Delta C_6^{( r )}}{R^6 } +| b |^2 \hbar\delta
   =  - \frac{C_6}{R^6} +| b |^2 \hbar\delta ,
\end{equation}
where $b=\Omega/\left({2\delta}\right)$ is the mixing parameter defined before. 
In Eq. (\ref{BOlr}), the first term is a change of the effective $C_6$ due to
Rydberg-dressing, which agrees with the result discussed before, and the
second term represents a shift of the collision threshold (absent in our previous
discussion).
At shorter distance where $\Delta C_6^{( r )} /R^6 \gg \hbar|\delta|$, the 
term $\Delta(R)$ becomes negligible compared to $C_6^{(g)}/R^6$
in Eq. (\ref{BO2b}) as it smoothly goes to zero; only the long-range 
part of the interaction is modified by the blue-detuned Rydberg-dressing field. 
The transition between the two regimes is shown in Fig.~\ref{fig:sketch-pot}(c).

Similarly, the curve between the Rydberg atom and the molecule
is dressed by the laser, effectively shifted down by $\Delta(R)$:
(with $| \Omega /\delta| \ll 1$ and $C_6^{\left( r \right)} > C_6^{\left( g \right)}$) 
\begin{equation}\label{BO2c}
   U_r ( R ) =  - \frac{C_6^{(r)}}{R^6} -\hbar\delta - \Delta ( R )\;,
\end{equation}
which, in the large $R$-limit ($\Delta C_6^{( r )} /R^6 \ll \hbar|\delta |$),  becomes 
\begin{equation}
    U_r ( R ) \to - \frac{C_6^{( r )}  - | b |^2 \Delta C_6^{( r )}}{R^6 } -(1+|b|^2)\hbar\delta \;.
\end{equation}
This impliesa slightly smaller effective van der Waals coefficient and slightly larger 
effective detuning.

To compute the effect of Rydberg-dressing on chemical reactions, we adopted
the H$_2$+D  electronic ground PES of
Ref.~\cite{BKMPJCP1996}, already tested at ultracold temperatures \cite{SimbotinPCCP2011}.
The potential $V({\bf r})\equiv V(r_{12},r_{23},r_{31})$ 
depends on the internuclear distances ${\bf r} \equiv \{r_{12}, r_{23}, r_{13}\}$, 
where 1 stands for D, and 2 and 3 for the two identical H atoms. The distances between H$_2$ and D then is simply given 
as $R = \frac{1}{2} \sqrt {2\left( {r_{12}^2  + r_{31}^2 } \right) - r_{23}^2 }$. 
Similar to Eq. (\ref{BO2b}), the three-body surface of H$_2$ interacting with a
blue-detuned Rydberg-dressed D atom can be written as
\begin{equation}
   \tilde{V} ( {\bf r}) = V( {\bf r}) + \Delta (R) \;,
\end{equation}
where the blue-detuning ensure the smooth transition from the dressed
long-range PES to the bare PES at shorter $R$ without incurring possible
avoided-crossing for red-detunings \cite{RolstonPRA2010}. We choose 
$b$ so that the transition occurs far enough from the van der Waals complex
well (minimum of the blue curve in Fig.~\ref{fig:sketch-pot}).
We obtain the $S$-matrix by performing a fully quantum mechanic scattering calculation
using the ABC code of Manolopoulos and coworkers \cite{ABC2000} modified
for the ultracold regime \cite{SimbotinPCCP2011, PRL-H2Cl}.
The state-to-state cross sections are given by
\begin{equation}
\sigma _{q' \leftarrow q}^J \left( E \right) = \frac{\pi }{{k_q^2 \left( {2j + 1} \right)}}\sum\limits_{\ell,\ell'} {\left| {\delta _{q'q}  - S_{q'\ell'q\ell}^{J} \left( E \right)} \right|^2 }, 
\end{equation}
where $q=\{a,v,j\}$ is the set quantum numbers 
describing the molecular state (vibration $v$, rotation $j$, and arrangement $a$):
$a$ distinguishes the final state H$_2$+D (quenching) 
from HD+H (reaction). $J$ is the three-body total angular quantum 
number, and $\ell$ indicates the relative 
angular momentum between the initial reactants H$_2$ and D. The scattering 
wave number $k_q$ is defined by
\begin{equation}
    \hbar^2 k_q^2 \equiv E_c  = 2\mu _{\rm H_2  + D} 
   \left[ E - \left( \varepsilon _{v,j}  + | b |^2 \hbar\delta  \right) \right] \;,
\end{equation}
where $E_c$ is the collisional energy and $\varepsilon _{v,j}$ is the initial rovibrational energy of H$_2$ (in state $\{v,j\}$) ,
$| b|^2 \hbar\delta$ corresponds to 
the shift due to the Rydberg dressing 
shown in Eq. (\ref{BOlr}), and $\mu^{-1}_{\rm{H_2  + D}}= m^{-1}_{\rm H_2}+m^{-1}_{\rm D}$
is the reduced mass. We define the total energy-dependent inelastic rate as
\begin{equation}
   \kappa _{\rm{in}}( E ) = v_{\rm rel} \sum\limits_{J,q'\neq q} ( 2J + 1) \sigma _{q' \leftarrow q}^{J} ( E ) \; ,
\end{equation}
where $v_{\rm rel}=\hbar k_q/\mu_{\rm{H_2  + D}}$ is the relative velocity . 
The sum is over all final channels but the entrance channel.
Quenching/reaction rates ($\kappa _{\rm{Q/R}}$) are obtained
by splitting the sum with $a'=a$ and $a'\neq a$, respectively \cite{SimbotinPCCP2011}.

\begin{figure}[t]
\includegraphics[width=0.45 \textwidth]{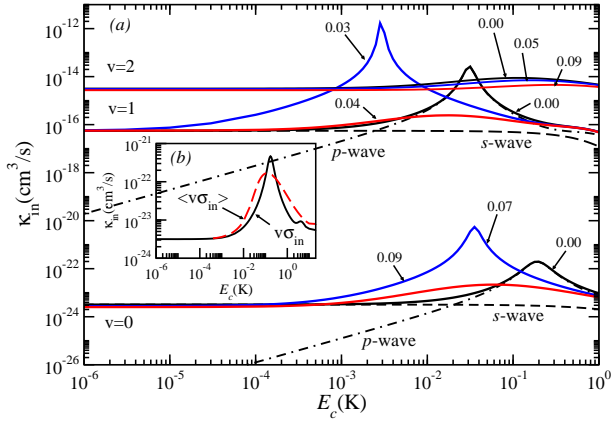}
\caption{(Color online) (a) $\kappa_{\rm in}$ vs. $E_c$ for different Rydberg 
mixing $\left|b\right|$ (given for each curve), for H$_2$($v=0,1,2,j=0$)+D. 
For the bare case ($|b|=0$) of $v=0$ and 1,
the $s$-wave (dashed line) and $p$-wave (dot-dashed line) components are shown. 
(b) Comparison of $\kappa_{\rm in}$ and its thermal average.}\label{Rates}
\end{figure}

Numerical results for $\kappa _{\rm{in}}$ in the ``bare"
case ($|b|=0$) for H$_2$($v=0,1,2$, $j=0$)+D are
shown in Fig.~\ref{Rates}(a). Resonances 
occur for both $v=0$ and 1, but not 2; the $s$-wave 
($\ell=0$) and $p$-wave ($\ell=1$) components are also shown, revealing
the $p$-wave nature of these resonances. By
varying the amount of Rydberg-dressing $|b|$,
the resonances can be moved substantially, while the non-resonant $v=2$
is only slightly affected. In these calculations, we fix $\delta \approx 296.09 \times 2 \pi$ GHz, 
so that the  threshold shift $|b|^2\hbar\delta$ of $U_g$ affects the position of the 
bound state in the entrance arrangement of the van der Waals complex. 
Even a modest $|b|$ is sufficient to
move a resonance significantly (cases $v=0$ and 1), while much larger values
are required if the van der Waals complex is not just about to support a
new bound-state ($v=2$).  Thus we limit our investigation to values of
$|b|=|\Omega/(2\delta)|\leq 0.1$, implying a laser intensity less than
$3\times10^8$W/cm$^2$. 
The $v=1$ level is particularly sensitive to a weak amount of Rydberg-dressing, with the
resonance having a larger magnitude and moving to much lower energy for $|b|=0.03$,
and simply disappearing for a slightly larger $|b|\approx 0.04$, 
when the quasi-bound state in the van der Waals complex becomes
bound. Figure~\ref{Rates}(b) compares the bare $\kappa _{\rm{in}}$ ($v=0$) with its thermal 
average (using a Maxwell distribution of $v_{\rm rel}$ characterized by a temperature $T$):
the agreement between both will become better as the resonance moves to lower $E_c$.

\begin{figure}[b]\vspace{-.65in}
\includegraphics[width=0.45 \textwidth]{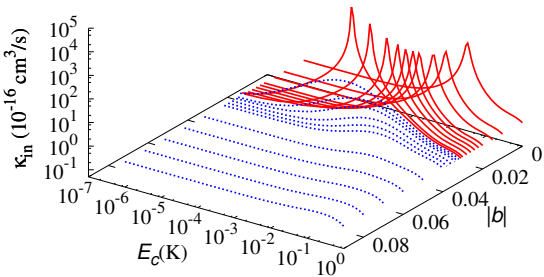} \vspace{-.15in}
\caption{(Color online) $\kappa_{\rm in}$ of D+H$_2$$(v=1,j=0)$ as a function of 
$\left|b\right|$ and $E_c$. The solid red curves show the 
inelastic rate with resonances when the H$_2 \cdots$D complex is not yet 
bound, and the dashed blue curves show the case when the complex 
becomes bound. Both $E_c$ and $\kappa_{\rm in}$ are on a logarithmic scale.}
\label{ratev1l0}
\end{figure}

\begin{figure}[t]
\includegraphics[width=0.45 \textwidth]{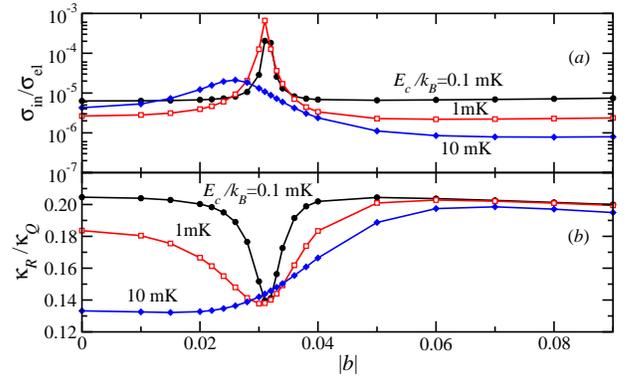}
\caption{(Color online) Ratios between inelastic and elastic cross sections (a), and
  between reaction and total inelastic rates (b), for three energies
  as a function of $|b|$.}
\label{Ratio}
\end{figure}

Figure~\ref{ratev1l0} shows this sensitivity of $\kappa_{\rm in}$ 
($v=1$) by varying $|b|$ and $E_c$;  as $|b|$ increases, the resonance
shifts to lower $E_c$ with an increased magnitude until it dissapears near $|b|\approx 0.04$, 
at which point the van der Waals complex acquires a new bound-state. As $|b|$
increases still, the maximum in $\kappa _{\rm in}$ starts shifting to
larger $E_c$ with a decreasing magnitude.  This example shows that 
one can, with modest Rydberg-dressing,  adjust and control 
$\kappa _{\rm in}$ in chemically active systems, by not only moving the position of resonances but also increasing their magnitude
by several orders. For example, if $E_c$ (or temperature) of a given experiment is near the 
resonance, the rate could be
reduced by moving the resonance away, or in cases where $E_c$
is not near the resonance, Rydberg-dressing
could move it to the right energy range. The ratio of inelastic to elastic cross sections in Fig.~\ref{Ratio}(a) shows
the sharp increase (near $|b|\approx 0.03$) in relative inelasticity. Note that elastic 
processes are less relevant from a chemical perspective since reactants
stay in their initial states. The branching ratio $\kappa_R/\kappa_{\rm in}$ for $v=1$
in Fig.~\ref{Ratio}(b) shows how chemical reactions can be controlled within a factor of 
two by varying $|b|$ near 0.03; for $v=0$, where only reaction channels exist 
 \cite{SimbotinPCCP2011}, the reaction rate can be changed by several orders of magnitudes
 (see Fig.~\ref{Rates}). Fig.~(\ref{Ratio}) suggests that the ratios 
 $\sigma_{\rm in}/\sigma_{\rm el}$ and $\kappa_R/\kappa_{\rm in}$ can be modified
 and controlled by Rydberg dressing.

In conclusion, we have shown that ultracold chemical reactions can be manipulated and 
modified using Rydberg-dressing. Although the concept 
was illustrated by dressing an atom approaching a diatomic molecule using the
benchmark H$_2$+D system,  this approach is general and could be applied to a wide range of systems. If the molecule
has a permanent dipole moment, {\it e.g.} for the system LiH+H, Rydberg-dressing the 
atom would lead  to a long-range interaction dominated by strong dipole induced 
$C_6$ due to the large dipole of LiH. With additional external electric fields, strong 
dipole-dipole interaction will become important. These strong long-range interaction 
might eventually lead to different branch ratios into exit channels. We note that Rydberg-dressing the final products instead of the reactants might allow one to direct the flux of probability into specific channels, and thus to
control the branching ratios for different final channels. This would open up
the possibility of state-to-state control of chemical reactions.

This work was partially supported by the US Department of Energy,
Office of Basic Energy Sciences (JW), the Air Force Office of
Scientific Research MURI award FA9550-09-1-0588 (JB), the Army Research Office 
Chemistry Division (IS), 
and the National Science Foundation
Grant No. PHY 1101254 (RC).

\bibliographystyle{apsrev}
\bibliography{refs}

\begin{thebibliography}{47}
\bibitem[{\citenamefont{Simbotin et~al.}(2011)\citenamefont{Simbotin, Ghosal,
  and C\^{o}t\'{e}}}]{SimbotinPCCP2011}
\bibinfo{author}{\bibfnamefont{I.}~\bibnamefont{Simbotin}},
  \bibinfo{author}{\bibfnamefont{S.}~\bibnamefont{Ghosal}}, \bibnamefont{and}
  \bibinfo{author}{\bibfnamefont{R.}~\bibnamefont{C\^{o}t\'{e}}},
  \bibinfo{journal}{Phys. Chem. Chem. Phys.} \textbf{\bibinfo{volume}{13}},
  \bibinfo{pages}{19148} (\bibinfo{year}{2011}), \bibinfo{note}{we follow the
  notations in this reference to denote any process that is not elastic as
  inelastic process, including the chemical reaction process where the final
  products are different from the initial reactants and the quenching process
  where the final products are the same as the initial reactants.}
\end{thebibliography}

\end{document}